\documentclass[amsmath,amssymb,pra,aps,showpieces,superscriptaddress,twocolumn]{revtex4-2}
\usepackage{amsfonts,amsthm,graphics,graphicx,epsfig,bbm}
\usepackage[colorlinks=true,citecolor=blue,linkcolor=blue,urlcolor=blue]{hyperref}
\usepackage[usenames]{color}
\usepackage{graphicx}
\usepackage{subfigure}
\usepackage{epsfig}
\usepackage{dcolumn}
\usepackage{bm}
\usepackage{color}
\usepackage{times}
\usepackage{epstopdf}
\usepackage{amstext}
\usepackage{latexsym}
\usepackage{amsfonts}
\usepackage{psfrag}
\usepackage{soul,xcolor}
\usepackage[normalem]{ulem}
\usepackage{dsfont}
\usepackage{txfonts}
\usepackage{float}
\usepackage{algorithm}
\usepackage{algpseudocode}

\newcommand{\ket}[1]{\vert #1 \rangle}
\newcommand{\bra}[1]{\langle #1 \vert}
\newcommand{\ketbra}[2]{\vert #1 \rangle \langle #2 \vert}
\newcommand{\braket}[2]{\langle #1 \vert #2 \rangle}



\usepackage{tikz,xcolor}
\definecolor{lime}{HTML}{A6CE39}
\DeclareRobustCommand{\orcidicon}{%
	\begin{tikzpicture}
	\draw[lime, fill=lime] (0,0) 
	circle [radius=0.16] 
	node[white] {{\fontfamily{qag}\selectfont \tiny ID}};
	\draw[white, fill=white] (-0.0625,0.095) 
	circle [radius=0.007];
	\end{tikzpicture}
	\hspace{-2mm}
}

\foreach \x in {A, ..., Z}{%
	\expandafter\xdef\csname orcid\x\endcsname{\noexpand\href{https://orcid.org/\csname orcidauthor\x\endcsname}{\noexpand\orcidicon}}
}

\begin{document}

\setstcolor{red}

\title{Role of spectral structure in adiabatic ground-state preparation of the XXZ model} 
\date{\today}

\author{Francisco Albarr\'an-Arriagada \orcidA{}} 

\affiliation{Kipu Quantum GmbH, Greifswalderstrasse 212, 10405 Berlin, Germany}
\affiliation{Departamento de F\'isica, CEDENNA, Universidad de Santiago de Chile (USACH), Avenida V\'ictor Jara 3493, 9170124, Santiago, Chile.}
\email[F. Albarr\'an-Arriagada]{\qquad francisco.albarran@usach.cl}

\author{Juan Carlos Retamal \orcidB{}}
\affiliation{Departamento de F\'isica, CEDENNA, Universidad de Santiago de Chile (USACH), Avenida V\'ictor Jara 3493, 9170124, Santiago, Chile.}
\email[J. C. Retamal]{\qquad juan.retamal@usach.cl}

\date{\today}

\begin{abstract}
Adiabatic ground-state preparation is fundamentally limited by the spectral structure of the time-dependent Hamiltonian, particularly by gap reductions and degeneracies that induce nonadiabatic transitions. We examine this dependence in the anisotropic Heisenberg (XXZ) model on an eight-site ring by comparing three strategies: optimization of the initial Hamiltonian, addition of auxiliary terms, and considering approximate counterdiabatic driving. Owing to anisotropy-dependent level crossings among low-energy states, the XXZ model provides a stringent benchmark. We find that performance is mainly constrained by spectral degeneracies between the ground and excited states. Simple strategies such as initial-Hamiltonian optimization or site-dependent Zeeman fields, suppresses critical crossings and drastically enhance ground-state preparation. In contrast, counterdiabatic terms alone do not improve the protocol when the spectral structure remains level-crossings, becoming effective only after degeneracies are removed. These results identify spectral engineering as a prerequisite for efficient adiabatic ground-state preparation in interacting spin systems.
\end{abstract}

\maketitle

\section{Introduction}

Preparing the ground state of interacting quantum systems is a central problem in quantum computing and quantum information science. In quantum computing, cost functions can be encoded into Hamiltonians whose ground states provide the solution to optimization problems, such as in QUBO and HUBO formulations~\cite{Bharti2022RevModPhys,Abbas2024NatRevPhys}. In quantum simulation, obtaining the ground state of strongly correlated Hamiltonians arising in condensed matter physics or quantum chemistry is essential for understanding their physical properties~\cite{Georgescu2014RevModPhys,Daley2022Nature}. 

Adiabatic quantum evolution offers a conceptually simple framework to address this task: the system is initialized in the ground state of an easily preparable Hamiltonian and slowly evolved to a target Hamiltonian whose ground state encodes the state of interest~\cite{Barends2016Nature,Albash2018RevModPhys}. According to the quantum adiabatic theorem, the system follows the instantaneous ground state provided that the evolution is sufficiently slow and the spectrum remains nondegenerate along the interpolation. In practice, however, the performance of adiabatic protocols is strongly constrained by the spectral structure of the time-dependent Hamiltonian. Gap reductions, avoided crossings, and degeneracies between low-lying states induce nonadiabatic transitions that severely limit the fidelity of ground-state preparation~\cite{GarciaPintos2023PhysRevLett}.

To mitigate these limitations, several strategies have been proposed to accelerate quantum evolutions while preserving high fidelity. Shortcuts to adiabaticity (STA), including counterdiabatic driving~\cite{Chen2010PhysRevLett,GueryOdelin2019RevModPhys,DelCampo2013PhysRevLett,Nakahara2022PTRSA}, introduce additional terms designed to suppress diabatic transitions. These methods have demonstrated significant potential in both digital~\cite{Hegade2021PhysRevAppl} and analog settings~\cite{Wang2019PhysRevAppl,Zhou2020PhysRevAppl,Zhang2024arXiv}. Other approaches include fast-forward scaling theory~\cite{Masuda2022PTRSA}, time-rescaling protocols~\cite{Bernardo2020PhysRevRes}, variational optimization of driving Hamiltonians~\cite{Sels2017PNAS}, and schedule optimization~\cite{Finzgar2024PhysRevRes}. While these techniques aim to suppress nonadiabatic excitations, their effectiveness ultimately depends on the underlying spectral properties of the system. In particular, when the interpolation exhibits degeneracies or changes in the ordering of low-energy levels, transitionless driving alone may not ensure successful ground-state preparation.

These observations motivate a systematic investigation of how controlled modifications of the spectral structure impact adiabatic performance. Rather than attempting full spectral engineering, which is generally intractable in many-body systems, we focus on two physically intuitive modifications. First, we introduce auxiliary site-dependent Zeeman terms, which are known to lift degeneracies and induce level splittings in interacting spin systems. Second, we consider optimizing the initial Hamiltonian through local rotations, preparing a separable initial state that is energetically closer to the target ground state while preserving experimental simplicity. Both strategies modify the spectrum along the interpolation while leaving the initial and final Hamiltonians unchanged. Later, we also consider the addition of counterdiabatic driving, studying the role of the spectral modification in the performance of this STA technique.

As a testbed, we study the anisotropic Heisenberg (XXZ) model on an eight-site ring, which exhibits anisotropy-dependent level crossings, changes in the ordering of low-energy states, and strong correlations~\cite{Gu2003PhysRevA}. These features make it a stringent benchmark for adiabatic ground-state preparation. By comparing standard adiabatic evolution with the addition of auxiliary fields, initial Hamiltonian optimization, and approximate counterdiabatic driving, we clearly show the role of spectral modification in transition suppression. We show that removing detrimental degeneracies is a prerequisite for the effectiveness of counterdiabatic corrections, establishing the spectral structure as the central factor governing adiabatic ground-state preparation in interacting spin systems.

\section{Model and Adiabatic Protocol}

\begin{figure*}[t]
\centering
	\includegraphics[width=1\linewidth]{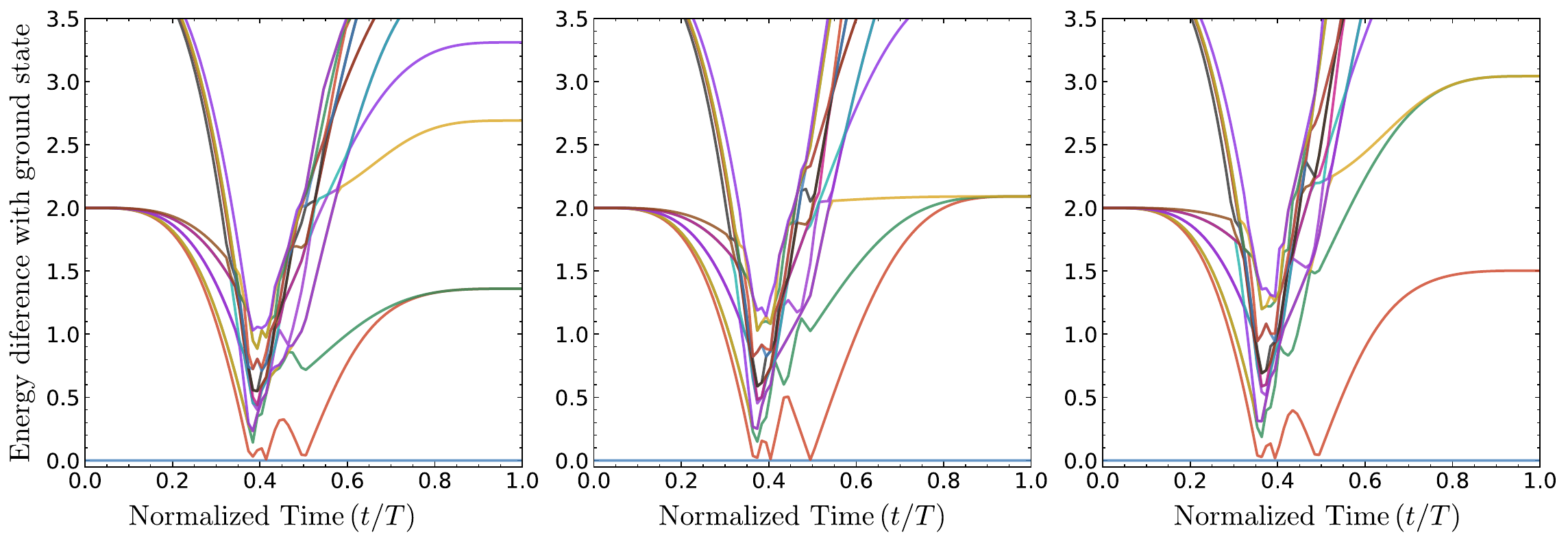}
	\caption{Energy spectrum as a function of normalized time for pure adiabatic evolution. Left panel $\Delta=0.5$. Central panel $\Delta=1$, Right panel $\Delta=1.5$}
	\label{Fig01}
\end{figure*}

\subsection{Adiabatic interpolation}
We consider a one-dimensional spin-$1/2$ ring with anisotropic Heisenberg ($XXZ$) interactions described by
\begin{equation}
H_f = J \sum_{j=1}^{n} 
\left(
\sigma_j^x \sigma_{j+1}^x 
+ \sigma_j^y \sigma_{j+1}^y 
+ \Delta\, \sigma_j^z \sigma_{j+1}^z
\right),
\label{Eq01}
\end{equation}
where $\sigma_j^{\alpha}$ ($\alpha = x,y,z$) are Pauli operators acting on site $j$, $J$ is the exchange coupling constant, and periodic boundary conditions are imposed by identifying $\sigma_{n+1}^{\alpha} \equiv \sigma_{1}^{\alpha}$. The anisotropy parameter $\Delta$ controls the relative strength between longitudinal and transverse couplings.

The XXZ chain exhibits a rich phase diagram in the thermodynamic limit, including a quantum phase transition at $\Delta=-1$ separating ferromagnetic and correlated phases~\cite{Gu2005PhysRevA}. Even for finite system sizes, the model retains nontrivial correlation properties and frustration effects for an even number of sites~\cite{Kubo1988PhysRevLett}. In addition, the low-energy spectrum displays anisotropy-dependent level crossings and changes in the ordering of eigenstates, making the finite XXZ ring a stringent setting to analyze adiabatic ground-state preparation.

In this work we consider $n=8$ sites and focus on $\Delta \in \{0.5,\,1.0,\,1.5\}$, spanning regimes around the isotropic point. The dense low-energy spectrum and the presence of crossings between low-lying states provide a suitable benchmark to investigate the role of spectral structure in adiabatic evolution.

\begin{figure}[t]
\centering
	\includegraphics[width=0.9\linewidth]{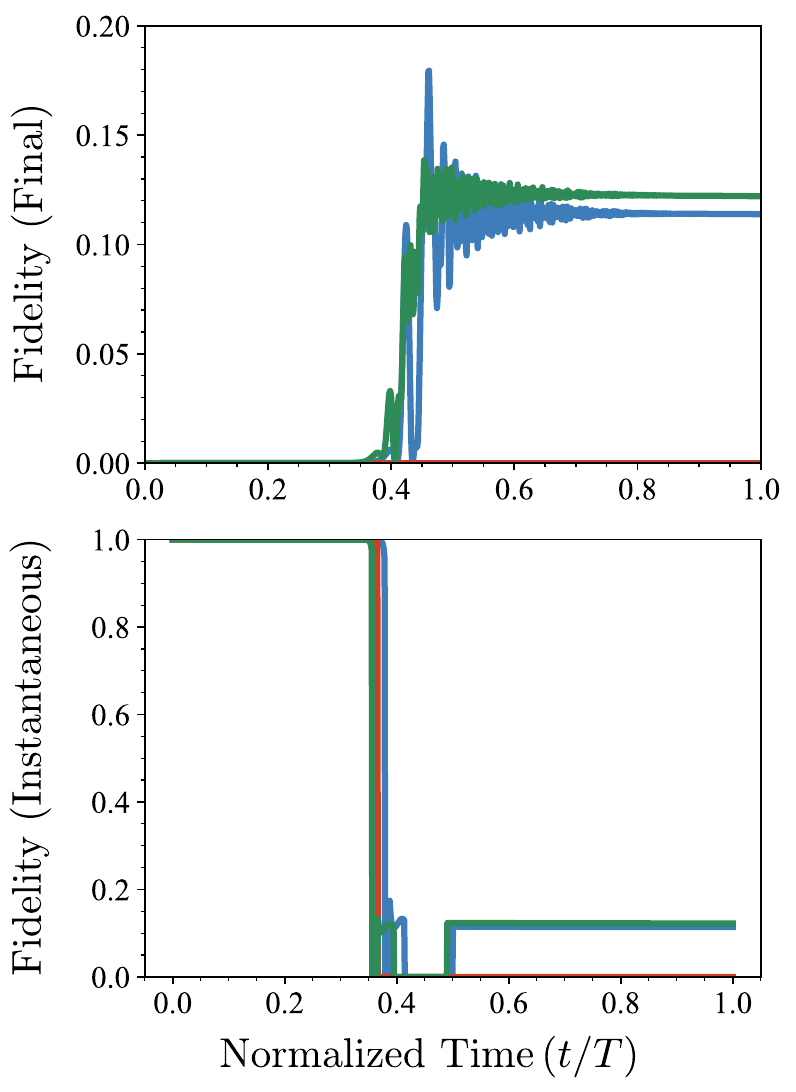}
	\caption{Standard adiabatic evolution performance to reach the ground state for the XXZ model considering total evolution time $T=300J^{-1}$, $\Delta=0.5$ (blue line), $\Delta=1.0$ (red line) and $\Delta=1.5$ (green line). Top panel: Fidelity between the state at time $t$ against the ground state of final Hamiltonian $H_f$. Bottom panel: Fidelity between the state at time $t$ and the instantaneous ground state of $H(t/T)$, see Eq. (\ref{Eq02}).}
	\label{Fig02}
\end{figure}

\subsection{Standard adiabatic evolution}

Adiabatic ground-state preparation is implemented through a continuous interpolation between an initial Hamiltonian $H_i$, whose ground state is easily prepared, and the target Hamiltonian $H_f$ defined in Eq.~(\ref{Eq01}). The time-dependent Hamiltonian is reads
\begin{equation}
H_{\mathrm{ad}}(t) = [1-\lambda(t/T)] H_i + \lambda(t/T) H_f,
\label{Eq02}
\end{equation}
where $T$ denotes the total evolution time and $\lambda(s)$, with $s=t/T \in [0,1]$, is a smooth scheduling function satisfying $\lambda(0)=0$ and $\lambda(1)=1$. Throughout this work we consider
\begin{equation}
\lambda(s)=\sin^2\!\left[\frac{\pi}{2}\sin^2\!\left(\frac{\pi s}{2}\right)\right],
\end{equation}
which ensures smooth boundary conditions and vanishing derivatives at the endpoints.

The most frequent choice for the initial Hamiltonian is a local interaction along the $x$-axis given by $H_i=-\epsilon\sum_j\sigma_j^x$, where $\epsilon$ is the initial free energy. The ground state of this Hamiltonian is a homogeneous superposition of all the states in the computational basis. Unless otherwise stated, we set $\epsilon = J = 1$.

In the limit $T \rightarrow \infty$, the adiabatic theorem guarantees that the system follows the instantaneous ground state of $H_{\mathrm{ad}}(t)$ provided that the spectrum remains nondegenerate and sufficiently gapped throughout the evolution. For finite $T$, transitions to excited states may occur, particularly in the presence of gap reductions or degeneracies between low-lying levels. The impact of such spectral features on the evolution is central to the present study.

This protocol will serve as a reference and will be denoted as standard adiabatic evolution (SA) in the following.

\begin{figure*}[t]
\centering
	\includegraphics[width=1\linewidth]{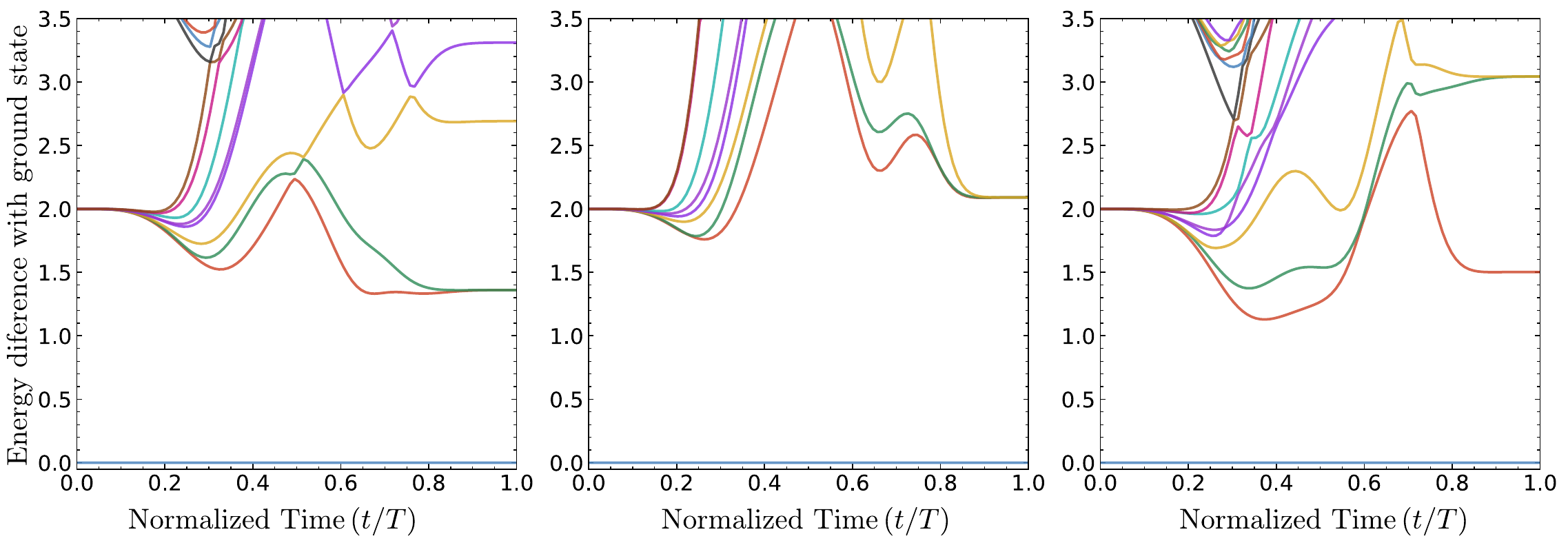}
	\caption{Energy spectrum as a function of normalized time for adiabatic evolution with auxiliary Hamiltonian. Left panel $\Delta=0.5$. Central panel $\Delta=1$, Right panel $\Delta=1.5$}
	\label{Fig03}
\end{figure*}

\subsection{Auxiliary Hamiltonian}

As our main goal is to avoid possible degenerancy points, our first strategy will be to add an auxiliary local Zeeman term in order to induce energy level splitting. We modify the adiabatic Hamiltonian as follows
\begin{equation}
H(t) = [1-\lambda(t/T)] H_i + \lambda(t/T) H_f 
+ \lambda(t/T)\big[1-\lambda(t/T)\big] H_{\mathrm{aux}},
\label{Eq07}
\end{equation}
so that $H(0)=H_i$ and $H(T)=H_f$. The auxiliary contribution, only changes the spectral structure along the evolution.

We consider a site-dependent longitudinal field (Zeeman term) of the form
\begin{equation}
H_{\mathrm{aux}} = \sum_{j=1}^{n} \omega_j \sigma_j^z,
\end{equation}
where $\omega_j$ are real parameters. As we mention before, such Zeeman terms are known to lift degeneracies and induce level splittings in interacting spin systems. The parameters $\{\omega_j\}$ are treated variationally and are determined by minimizing the expectation value of $H_f$ at the end of the evolution. This modification preserves locality while allowing controlled adjustments of the instantaneous spectrum, being experimentally feasible. In this case the total number of variational parameters is $n$, that is a linearly with the number of qubits.

\subsection{Initial Hamiltonian optimization}

A second strategy, is the modification of the initial Hamiltonian. We generalize the initial Hamiltonian by allowing arbitrary local spin orientations while preserving its single-site structure, that is
\begin{equation}
H_i = \epsilon \sum_{j=1}^{n} \hat{u}_j \cdot \vec{\sigma}_j,
\label{Eq04}
\end{equation}
where $\vec{\sigma}_j = (\sigma_j^x, \sigma_j^y, \sigma_j^z)$ and 
\begin{equation}
\hat{u}_j =
(\sin\theta_j \cos\phi_j,\,
 \sin\theta_j \sin\phi_j,\,
 \cos\theta_j)
\end{equation}
is a unit vector parametrized by the angles $\theta_j$ and $\phi_j$, indicating the direcction of the spin in the site $j$ in the Bloch sphere.

Since $H_i$ remains a sum of local terms, its ground state is a separable product state of the form
\begin{equation}
\ket{\Phi(\{\theta_j\},\{\phi_j\})}
=
\bigotimes_{j=1}^{n}
\ket{\psi_j(\theta_j,\phi_j)}.
\end{equation}

The optimal parameters are obtained by minimizing the expectation value of the target Hamiltonian,
\begin{equation}
\min_{\{\theta_j\},\{\phi_j\}}
\,
\bra{\Phi(\{\theta_j\},\{\phi_j\})}
H_f
\ket{\Phi(\{\theta_j\},\{\phi_j\})},
\end{equation}
that is, an initial state that is energetically closer to the ground state of $H_f$ while retaining a simple local structure. With this, we modify the spectral properties along the interpolation without changing the locality of the initial Hamiltonian. We note that the total number of variational parameters in this case is $2n$, that is a linear scaling, but with a factor $2$ in comparison with the previous case.

\subsection{Counterdiabatic driving}

A third strategy consists in to add to the adiabatic Hamiltonian a counterdiabatic (CD) terms designed to suppress diabatic transitions~\cite{Berry2009JPhysA}. For a time-dependent Hamiltonian $H_{\mathrm{ad}}(t)$ with instantaneous eigenstates $\ket{n(t)}$ and eigenvalues $E_n(t)$ satisfying
\begin{equation}
H_{\mathrm{ad}}(t)\ket{n(t)} = E_n(t)\ket{n(t)},
\end{equation}
the exact counterdiabatic Hamiltonian reads
\begin{equation}
H_{\mathrm{cd}}(t)
=
i\hbar
\sum_{m\neq n}
\frac{
\ketbra{m(t)}{m(t)}
\partial_t H_{\mathrm{ad}}(t)
\ketbra{n(t)}{n(t)}
}{
E_n(t)-E_m(t)
}.
\label{Eq09}
\end{equation}
The full evolution is then governed by
\begin{equation}
H(t) = H_{\mathrm{ad}}(t) + H_{\mathrm{cd}}(t),
\end{equation}
which enforces transitionless dynamics in the absence of degeneracies.

The exact construction of $H_{\mathrm{cd}}(t)$ requires complete knowledge of the instantaneous spectrum and is generally intractable for interacting many-body systems. Therefore, We consider a first-order approximation based on the nested-commutator expansion~\cite{Claeys2019PhysRevLett}, given by
\begin{equation}
H_{\mathrm{cd}}(t)
\approx
i\,\dot{\lambda}(t)\,
\alpha
\left[
H_{\mathrm{ad}}(t),
\partial_{\lambda} H_{\mathrm{ad}}(t)
\right],
\label{Eq11}
\end{equation}
where $\alpha$ is treated as a real variational parameter. We impose $|\alpha| \le w$, with $w$ representing a constraint set by experimental feasibility, in our case we fix it in $w=10\epsilon$

Although analytical expressions for $\alpha$ is given in Ref.~\cite{Xie2022PhysRevB}, in this work it is determined variationally by minimizing the expectation value of the target Hamiltonian $H_f$ at the final time. This choice focuses the optimization on ground-state preparation rather than suppressing transitions across the entire spectrum.

\subsection{Performance metrics}

To quantify the efficiency of the different protocols, we consider two complementary figures of merit.

First, we evaluate the normalized energy distance
\begin{equation}
\mathcal{N}(T)
=
\frac{
\langle \phi(0) | H_f | \phi(0) \rangle
-
\langle \phi(T) | H_f | \phi(T) \rangle
}{
\langle \phi(0) | H_f | \phi(0) \rangle
-
E_F
},
\label{Eq03}
\end{equation}
where $\ket{\phi(t)}$ is the evolved state at time $t$ and $E_F$ denotes the ground-state energy of $H_f$. By construction, $\mathcal{N}(T)=1$ corresponds to perfect ground-state preparation, while $\mathcal{N}(T)=0$ indicates no improvement relative to the initial state.

Second, we define the adiabatic fidelity
\begin{equation}
\mathcal{F}_{\mathrm{ad}}
=
\frac{1}{T}
\int_0^T
\left|
\braket{\phi(t)}{\Psi(t)}
\right|
\, dt,
\end{equation}
where $\ket{\Psi(t)}$ is the instantaneous ground state of the adiabatic Hamiltonian. The quantity $\mathcal{F}_{\mathrm{ad}} \in [0,1]$ measures if the evolution is following the instantaneous ground state throughout the protocol. We note that $\mathcal{F}_{\mathrm{ad}}=0$ indicates complete absence of overlap with the instantaneous ground state during the evolution, whereas $\mathcal{F}_{\mathrm{ad}}=1$ corresponds to an evolution that follows the groundstate at any time.

\begin{figure*}[t]
\centering
	\includegraphics[width=1\linewidth]{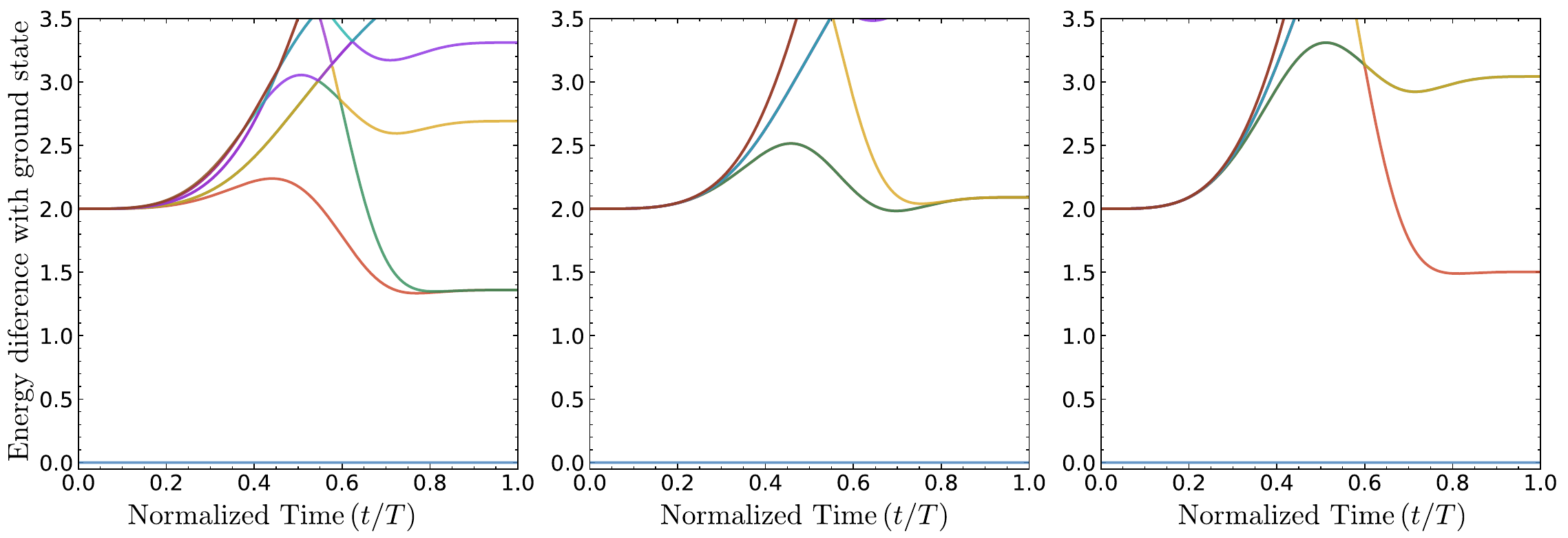}
	\caption{Energy spectrum as a function of normalized time for initial Hamiltonian optimization. Left panel $\Delta=0.5$. Central panel $\Delta=1$, Right panel $\Delta=1.5$}
	\label{Fig04}
\end{figure*}

\section{Results}

\begin{table}[b]
\begin{tabular}{|l|ll|ll|ll|}
\hline
 & $T=1$&& $T=3$&&  $T=10$&\\ \hline
 & \multicolumn{1}{l|}{$\mathcal{N}(T)$} & $\mathcal{F}_{ad}(T)$ & \multicolumn{1}{l|}{$\mathcal{N}(T)$} & $\mathcal{F}_{ad}(T)$ & \multicolumn{1}{l|}{$\mathcal{N}(T)$} & $\mathcal{F}_{ad}(T)$ \\ \hline
 S.A. &    \multicolumn{1}{l|}{0.01}  &      0.38&\multicolumn{1}{l|}{0.06}&0.39&  \multicolumn{1}{l|}{0.20}&0.40\\ \hline
  S.A. + A.H. & \multicolumn{1}{l|}{0.57}&0.39 &\multicolumn{1}{l|}{0.78} &0.50& \multicolumn{1}{l|}{0.94} &0.87\\ \hline
   S.A. + C.D.& \multicolumn{1}{l|}{0.56} & 0.55 & \multicolumn{1}{l|}{0.85} & 0.85 & \multicolumn{1}{l|}{1.00} & 0.99 \\ \hline
 O.I. & \multicolumn{1}{l|}{0.30} &   0.82  & \multicolumn{1}{l|}{0.88} & 0.94 & \multicolumn{1}{l|}{0.98} & 0.99   \\ \hline
 O.I. + A.H. & \multicolumn{1}{l|}{0.30}& 0.82 & \multicolumn{1}{l|}{0.88}& 0.94 & \multicolumn{1}{l|}{0.98}& 0.99 \\ \hline
 O.I. + C.D. & \multicolumn{1}{l|}{0.84}&0.94& \multicolumn{1}{l|}{0.92}&0.96& \multicolumn{1}{l|}{0.99}&0.96\\ \hline
O.I. + A.H. + C.D.& \multicolumn{1}{l|}{0.84}&0.94& \multicolumn{1}{l|}{0.92}&0.96& \multicolumn{1}{l|}{0.99}&0.96\\ \hline
\end{tabular}
\label{Tab01}
\caption{Results for $\Delta=0.5$ considering different strategies. S.A: Standard Adiabatic, A.H: Auxiliary Hamiltonian, O.I: Optimal Initial Hamiltonian. C.D: Counterdiabatic driving.}
\end{table}

\subsection{Spectral limitations of standard adiabatic evolution}

We begin by analyzing the reference standard adiabatic evolution (SA). Figure~\ref{Fig01} shows the instantaneous energy spectrum of $H_{\mathrm{ad}}(t)$ as a function of the normalized time $t/T$ for $\Delta=0.5$, $1.0$, and $1.5$. In all cases, the spectrum exhibits multiple crossings and near-degeneracies among low-lying states during the interpolation. In particular, crossings between the ground and first excited states occur in the intermediate stage of the evolution, signaling regions where the adiabatic condition is severely compromised.

The impact of these spectral features on the dynamics is illustrated in Fig.~\ref{Fig02}. The top panel shows the fidelity with respect to the ground state of $H_f$, while the bottom panel displays the overlap with the instantaneous ground state. Despite considering long evolution times ($T=300J^{-1}$), the final fidelity remains low for $\Delta=0.5$ and $\Delta=1.5$, and is strongly suppressed for the isotropic case $\Delta=1$. 

The adiabatic fidelity reveals pronounced drops precisely in the regions where the spectrum exhibits degeneracies or drastically gap reduction. These features induce diabatic transitions, allowing that the system populate excited states. Notably, in the isotropic case the change in the ordering of low-energy levels obstructs ground-state preparation even for large $T$. 

These observations indicate that the primary limitation of the SA protocol originates from the spectral structure of the interpolation, motivating the introduction of controlled modifications aimed at mitigating degeneracies and gap reductions.

\begin{table}[b]
\begin{tabular}{|l|ll|ll|ll|}
\hline
 & $T=1$                                 &                       & $T=3$                                 &                       &  $T=10$                                &                       \\ \hline
 & \multicolumn{1}{l|}{$\mathcal{N}(T)$} & $\mathcal{F}_{ad}(T)$ & \multicolumn{1}{l|}{$\mathcal{N}(T)$} & $\mathcal{F}_{ad}(T)$ & \multicolumn{1}{l|}{$\mathcal{N}(T)$} & $\mathcal{F}_{ad}(T)$ \\ \hline
 S.A. &    \multicolumn{1}{l|}{0.00}  &      0.37                &    \multicolumn{1}{l|}{0.00}      &   0.37                     &  \multicolumn{1}{l|}{0.00}   & 0.37                \\ \hline
  S.A. + A.H. & \multicolumn{1}{l|}{0.37} & 0.36 & \multicolumn{1}{l|}{0.62}  & 0.43 & \multicolumn{1}{l|}{0.90} & 0.71        \\ \hline
   S.A. + C.D.& \multicolumn{1}{l|}{0.00} & 0.37 & \multicolumn{1}{l|}{0.00} & 0.37 & \multicolumn{1}{l|}{0.00} & 0.37 \\ \hline
 O.I. & \multicolumn{1}{l|}{0.34} & 0.79 & \multicolumn{1}{l|}{0.92} & 0.95 & \multicolumn{1}{l|}{1.00} & 1.00     \\ \hline
 O.I. + A.H. & \multicolumn{1}{l|}{0.34}& 0.79 & \multicolumn{1}{l|}{0.92}& 0.95 & \multicolumn{1}{l|}{1.00}& 1.00 \\ \hline
 O.I. + C.D. & \multicolumn{1}{l|}{0.85}&0.95& \multicolumn{1}{l|}{0.95}&0.98& \multicolumn{1}{l|}{1.00}&0.94\\ \hline
O.I. + A.H. + C.D.& \multicolumn{1}{l|}{0.84}&0.94& \multicolumn{1}{l|}{0.95}&0.98& \multicolumn{1}{l|}{1.00}&0.94\\ \hline
\end{tabular}
\label{Tab02}
\caption{Results for $\Delta=1$ considering different strategies. S.A: Standard Adiabatic, A.H: Auxiliary Hamiltonian, O.I: Optimal Initial Hamiltonian. C.D: Counterdiabatic driving.}
\end{table}

\subsection{Effect of auxiliary fields}

We next examine the impact of the auxiliary Hamiltonian introduced in Eq.~(\ref{Eq07}). The instantaneous spectrum of the modified interpolation is shown in Fig.~\ref{Fig03} for $\Delta=0.5$, $1.0$, and $1.5$. 

The addition of site-dependent longitudinal fields alters the spectral structure during the evolution while preserving the initial and final Hamiltonians. In particular, the auxiliary term lifts several of the degeneracies observed in the standard adiabatic evolution, leading to an effective splitting of low-energy levels in the intermediate stages of the protocol. This behavior is consistent with the role of Zeeman fields in breaking spectral symmetries and inducing level separation in interacting spin systems.

Although the most pronounced crossings between the ground and first excited states are significantly reduced, residual near-degeneracies remain in parts of the spectrum. As a consequence, while the auxiliary modification mitigates the most detrimental spectral features, it does not completely eliminate the sources of diabatic transitions.

These results indicate that even simple local modifications can substantially reshape the instantaneous spectrum. However, further control over the spectral structure may be required to fully suppress nonadiabatic effects.

\subsection{Initial Hamiltonian optimization}

We now analyze the effect of optimizing the initial Hamiltonian while preserving its local structure. In the following, this protocol will be referred to as optimized initial Hamiltonian (OI). The instantaneous spectrum obtained under this modification is shown in Fig.~\ref{Fig04} for $\Delta=0.5$, $1.0$, and $1.5$.

Compared with the standard adiabatic evolution (SA), the OI protocol produces a substantial reorganization of the low-energy spectrum during the interpolation. In particular, several of the degeneracies and near-crossings between the ground and first excited states observed in Fig.~\ref{Fig01} are significantly reduced or completely removed. This modification results from preparing an initial separable state that is energetically closer to the target ground state, thereby smoothing the spectral landscape along the evolution.

Importantly, this strategy does not introduce additional interaction terms or increase the nonlocal character of the Hamiltonian. The initial Hamiltonian remains a sum of single-site operators, and the corresponding ground state retains a product-state structure. Nevertheless, the change in local spin orientations is sufficient to reshape the instantaneous spectrum in a nontrivial manner.

The OI protocol therefore achieves a pronounced suppression of detrimental crossings while maintaining structural simplicity, highlighting the sensitivity of adiabatic performance to the choice of initial Hamiltonian.

\subsection{Interplay with counterdiabatic driving}

We now examine the role of approximate counterdiabatic (CD) driving in combination with the different spectral modifications discussed above. The quantitative performance of all protocols is summarized in Tables~I--III for $\Delta=0.5$, $1.0$, and $1.5$, respectively.

When CD terms are added to the standard adiabatic evolution (SA+CD), only marginal improvement is observed in regimes where the instantaneous spectrum exhibits degeneracies or changes in the ordering of low-energy states. This behavior is particularly evident in the isotropic case $\Delta=1$ (Table~II), where both the normalized energy distance and the adiabatic fidelity remain essentially unchanged with respect to SA, even for increasing evolution times. In this regime, the spectral crossings prevent reliable adiabatic following, and the variational CD correction is unable to compensate for the unfavorable eigenstructure.

In contrast, when CD driving is applied after modifying the spectral structure, a substantial enhancement is obtained. For $\Delta=0.5$ and $\Delta=1.5$ (Tables~I and III), the combination OI+CD yields near-unit normalized energy distance and high adiabatic fidelity already at moderate evolution times. Most notably, in the isotropic case (Table~II), where SA and SA+CD fail to reach the target ground state, the optimized initial Hamiltonian (OI) alone produces a dramatic improvement, and its combination with CD further enhances performance at short times.

These results establish a clear improvement among the considered strategies. The dominant limitation of the adiabatic protocol originates from spectral degeneracies. Counterdiabatic driving becomes effective only after the instantaneous spectrum has been reshaped to remove critical crossings. Spectral modification, therefore, constitutes a prerequisite for efficient transition suppression in interacting spin systems.

\begin{table}[t]
\begin{tabular}{|l|ll|ll|ll|}
\hline
 & $T=1$                                 &                       & $T=3$                                 &                       &  $T=10$                                &                       \\ \hline
 & \multicolumn{1}{l|}{$\mathcal{N}(T)$} & $\mathcal{F}_{ad}(T)$ & \multicolumn{1}{l|}{$\mathcal{N}(T)$} & $\mathcal{F}_{ad}(T)$ & \multicolumn{1}{l|}{$\mathcal{N}(T)$} & $\mathcal{F}_{ad}(T)$ \\ \hline
 S.A. &    \multicolumn{1}{l|}{0.01}  &      0.36                &    \multicolumn{1}{l|}{0.04}      &   0.36                     &  \multicolumn{1}{l|}{0.12}   & 0.37                \\ \hline
  S.A. + A.H. & \multicolumn{1}{l|}{0.60} & 0.39 & \multicolumn{1}{l|}{0.56} & 0.47& \multicolumn{1}{l|}{0.94}  &0.82        \\ \hline
   S.A. + C.D.& \multicolumn{1}{l|}{0.58} & 0.58 & \multicolumn{1}{l|}{0.85} & 0.85 & \multicolumn{1}{l|}{1.00} & 0.99 \\ \hline
 O.I. & \multicolumn{1}{l|}{0.40}  & 0.84 & \multicolumn{1}{l|}{0.91} & 0.96 & \multicolumn{1}{l|}{0.99} & 0.99 \\ \hline
 O.I. + A.H. & \multicolumn{1}{l|}{0.84}& 0.94 & \multicolumn{1}{l|}{0.92}& 0.96 & \multicolumn{1}{l|}{1.00}& 0.99 \\ \hline
 O.I. + C.D. & \multicolumn{1}{l|}{0.84}&0.94& \multicolumn{1}{l|}{0.93}&0.97& \multicolumn{1}{l|}{0.99}&0.96\\ \hline
O.I. + A.H. + C.D.& \multicolumn{1}{l|}{0.84}&0.95& \multicolumn{1}{l|}{0.92}&0.97& \multicolumn{1}{l|}{1.00}&0.99\\ \hline
\end{tabular}
\label{Tab03}
\caption{Results for $\Delta=1.5$ considering different strategies. S.A: Standard Adiabatic, A.H: Auxiliary Hamiltonian, O.I: Optimal Initial Hamiltonian. C.D: Counterdiabatic driving.}
\end{table}

\subsection{Quantitative comparison of protocols and dynamical behavior}

A quantitative comparison of all protocols is reported in Tables~I--III for $\Delta=0.5$, $1.0$, and $1.5$, respectively, and for representative evolution times $T=1$, $3$, and $10$. In all anisotropy regimes, the standard adiabatic evolution (SA) yields poor performance, with both the normalized energy distance $\mathcal{N}(T)$ and the adiabatic fidelity $\mathcal{F}_{\mathrm{ad}}$ remaining low, consistent with the dense low-energy spectrum and the presence of crossings discussed above.

Adding auxiliary longitudinal fields (SA+AH) provides a clear improvement, indicating that lifting degeneracies through Zeeman-induced level splittings can substantially enhance adiabatic performance. This is particularly evident for $\Delta=0.5$ and $\Delta=1.5$ (Tables~I and~III), where SA+AH already achieves high $\mathcal{N}(T)$. Nevertheless, the improvement is not uniform across all regimes and does not fully resolve the limitations associated with unfavorable spectral ordering.

A more pronounced enhancement arises from optimizing the initial Hamiltonian (OI), which consistently increases both $\mathcal{N}(T)$ and $\mathcal{F}_{\mathrm{ad}}$ across all $\Delta$ values. Most notably, in the isotropic case $\Delta=1$ (Table~II), SA fails to prepare the ground state and SA+CD does not improve over SA, whereas OI yields near-unit performance already for $T=3$. This provides direct quantitative evidence that spectral modification is essential when crossings or low-energy reorderings obstruct adiabatic following.

Regarding the interplay between auxiliary fields and initial-state optimization, Tables~I--III show that OI+AH does not systematically outperform OI, indicating that once the dominant degeneracies are removed by one of the spectral-modification strategies, adding an additional local splitting term provides limited further benefit. In contrast, combining OI with counterdiabatic driving (OI+CD) systematically improves short-time performance and yields near-unit $\mathcal{N}(T)$ and high $\mathcal{F}_{\mathrm{ad}}$ already at $T=1$ and $T=3$ for $\Delta=0.5$ and $\Delta=1.5$, while also enhancing the isotropic case at short times (Table~II). This confirms that CD corrections become effective only after the instantaneous spectrum has been reshaped to remove the most detrimental crossings.

\begin{figure}[t]
\centering
	\includegraphics[width=0.9\linewidth]{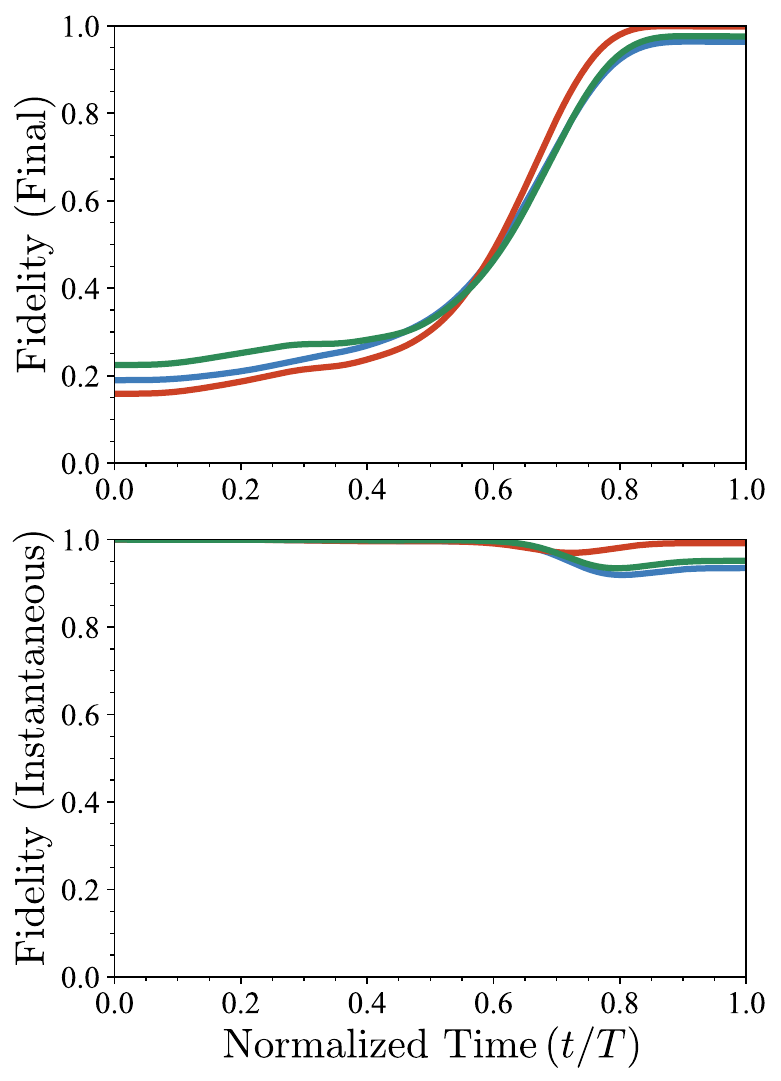}
	\caption{Optimil initial Hamiltonian with counterdiabatic driving strategy performance considering total evolution time $T=10J^{-1}$, $\Delta=0.5$ (blue line), $\Delta=1.0$ (red line) and $\Delta=1.5$ (green line). Top panel: Fidelity between the state at time $t$ against the ground state of final Hamiltonian $H_f$. Bottom panel: Fidelity between the state at time $t$ and the instantaneous ground state of $H(t/T)$, see Eq. (\ref{Eq02}).}
	\label{Fig05}
\end{figure}

The dynamical enhancement behind these strategies is illustrated in Fig.~\ref{Fig05}, where we show the time-resolved fidelities for the OI+CD protocol at $T=10$. In contrast to the SA dynamics, the evolution displays a smooth approach to the target ground state and a sustained overlap with the instantaneous ground state, without the pronounced oscillations associated with repeated diabatic transitions. Taken together, the numerical results establish that (i) spectral modification is the primary ingredient enabling efficient adiabatic ground-state preparation in this model, and (ii) counterdiabatic driving provides an additional, but secondary, improvement once the low-energy spectrum has been sufficiently regularized.

\section{Conclusion}

We have analyzed the role of spectral structure in adiabatic ground-state preparation of the finite XXZ spin chain by comparing standard adiabatic evolution, auxiliary fields, optimized initial Hamiltonians, and approximate counterdiabatic driving. Our results show that degeneracies and level crossings in the low-energy spectrum constitute the primary limitation of adiabatic protocols.

Simple local modifications, such as optimizing the initial Hamiltonian or introducing site-dependent Zeeman fields, substantially reshape the instantaneous spectrum and suppress detrimental crossings. In contrast, counterdiabatic driving alone does not overcome spectral obstructions and becomes effective only after the dominant degeneracies are removed. Among the considered strategies, optimization of the initial Hamiltonian emerges as the most impactful modification, achieving significant improvements while preserving locality and a separable initial state.

These findings establish spectral modification, particularly through initial-state optimization, as a prerequisite for efficient adiabatic ground-state preparation in interacting spin systems and provide guiding principles for the design of quantum control protocols aimed at accelerating adiabatic state preparation in many-body and condensed-matter quantum simulations.

\section{Acknowledgments}

We acknowledge financial support from Agencia Nacional de Investigaci\'on y Desarrollo (ANID): Financiamiento Basal para Centros Cient\'ificos y Tecnol\'ogicos de Excelencia grant No. AFB220001, Fondecyt grant 1231172. Also, the financial support of  Universidad de Santiago de Chile: DICYT Asociativo Grant No. 042431AA$\_$DAS.

\bibliography{References}

\begin{thebibliography}{25}%
\makeatletter
\providecommand \@ifxundefined [1]{%
 \@ifx{#1\undefined}
}%
\providecommand \@ifnum [1]{%
 \ifnum #1\expandafter \@firstoftwo
 \else \expandafter \@secondoftwo
 \fi
}%
\providecommand \@ifx [1]{%
 \ifx #1\expandafter \@firstoftwo
 \else \expandafter \@secondoftwo
 \fi
}%
\providecommand \natexlab [1]{#1}%
\providecommand \enquote  [1]{``#1''}%
\providecommand \bibnamefont  [1]{#1}%
\providecommand \bibfnamefont [1]{#1}%
\providecommand \citenamefont [1]{#1}%
\providecommand \href@noop [0]{\@secondoftwo}%
\providecommand \href [0]{\begingroup \@sanitize@url \@href}%
\providecommand \@href[1]{\@@startlink{#1}\@@href}%
\providecommand \@@href[1]{\endgroup#1\@@endlink}%
\providecommand \@sanitize@url [0]{\catcode `\\12\catcode `\$12\catcode
  `\&12\catcode `\#12\catcode `\^12\catcode `\_12\catcode `\%12\relax}%
\providecommand \@@startlink[1]{}%
\providecommand \@@endlink[0]{}%
\providecommand \url  [0]{\begingroup\@sanitize@url \@url }%
\providecommand \@url [1]{\endgroup\@href {#1}{\urlprefix }}%
\providecommand \urlprefix  [0]{URL }%
\providecommand \Eprint [0]{\href }%
\providecommand \doibase [0]{https://doi.org/}%
\providecommand \selectlanguage [0]{\@gobble}%
\providecommand \bibinfo  [0]{\@secondoftwo}%
\providecommand \bibfield  [0]{\@secondoftwo}%
\providecommand \translation [1]{[#1]}%
\providecommand \BibitemOpen [0]{}%
\providecommand \bibitemStop [0]{}%
\providecommand \bibitemNoStop [0]{.\EOS\space}%
\providecommand \EOS [0]{\spacefactor3000\relax}%
\providecommand \BibitemShut  [1]{\csname bibitem#1\endcsname}%
\let\auto@bib@innerbib\@empty
\bibitem [{\citenamefont {Bharti}\ \emph {et~al.}(2022)\citenamefont {Bharti},
  \citenamefont {Cervera-Lierta}, \citenamefont {Kyaw}, \citenamefont {Haug},
  \citenamefont {Alperin-Lea}, \citenamefont {Anand}, \citenamefont {Degroote},
  \citenamefont {Heimonen}, \citenamefont {Kottmann}, \citenamefont {Menke},
  \citenamefont {Mok}, \citenamefont {Sim}, \citenamefont {Kwek},\ and\
  \citenamefont {Aspuru-Guzik}}]{Bharti2022RevModPhys}%
  \BibitemOpen
  \bibfield  {author} {\bibinfo {author} {\bibfnamefont {K.}~\bibnamefont
  {Bharti}}, \bibinfo {author} {\bibfnamefont {A.}~\bibnamefont
  {Cervera-Lierta}}, \bibinfo {author} {\bibfnamefont {T.~H.}\ \bibnamefont
  {Kyaw}}, \bibinfo {author} {\bibfnamefont {T.}~\bibnamefont {Haug}}, \bibinfo
  {author} {\bibfnamefont {S.}~\bibnamefont {Alperin-Lea}}, \bibinfo {author}
  {\bibfnamefont {A.}~\bibnamefont {Anand}}, \bibinfo {author} {\bibfnamefont
  {M.}~\bibnamefont {Degroote}}, \bibinfo {author} {\bibfnamefont
  {H.}~\bibnamefont {Heimonen}}, \bibinfo {author} {\bibfnamefont {J.~S.}\
  \bibnamefont {Kottmann}}, \bibinfo {author} {\bibfnamefont {T.}~\bibnamefont
  {Menke}}, \bibinfo {author} {\bibfnamefont {W.-K.}\ \bibnamefont {Mok}},
  \bibinfo {author} {\bibfnamefont {S.}~\bibnamefont {Sim}}, \bibinfo {author}
  {\bibfnamefont {L.-C.}\ \bibnamefont {Kwek}},\ and\ \bibinfo {author}
  {\bibfnamefont {A.}~\bibnamefont {Aspuru-Guzik}},\ }\bibfield  {title}
  {\bibinfo {title} {Noisy intermediate-scale quantum algorithms},\ }\href
  {https://doi.org/10.1103/RevModPhys.94.015004} {\bibfield  {journal}
  {\bibinfo  {journal} {Rev. Mod. Phys.}\ }\textbf {\bibinfo {volume} {94}},\
  \bibinfo {pages} {015004} (\bibinfo {year} {2022})}\BibitemShut {NoStop}%
\bibitem [{\citenamefont {Abbas}\ \emph {et~al.}(2024)\citenamefont {Abbas},
  \citenamefont {Ambainis}, \citenamefont {Augustino}, \citenamefont
  {B{\"a}rtschi}, \citenamefont {Buhrman}, \citenamefont {Coffrin},
  \citenamefont {Cortiana}, \citenamefont {Dunjko}, \citenamefont {Egger},
  \citenamefont {Elmegreen}, \citenamefont {Franco}, \citenamefont {Fratini},
  \citenamefont {Fuller}, \citenamefont {Gacon}, \citenamefont {Gonciulea},
  \citenamefont {Gribling}, \citenamefont {Gupta}, \citenamefont {Hadfield},
  \citenamefont {Heese}, \citenamefont {Kircher}, \citenamefont {Kleinert},
  \citenamefont {Koch}, \citenamefont {Korpas}, \citenamefont {Lenk},
  \citenamefont {Marecek}, \citenamefont {Markov}, \citenamefont {Mazzola},
  \citenamefont {Mensa}, \citenamefont {Mohseni}, \citenamefont {Nannicini},
  \citenamefont {O'Meara}, \citenamefont {Tapia}, \citenamefont {Pokutta},
  \citenamefont {Proissl}, \citenamefont {Rebentrost}, \citenamefont {Sahin},
  \citenamefont {Symons}, \citenamefont {Tornow}, \citenamefont {Valls},
  \citenamefont {Woerner}, \citenamefont {Wolf-Bauwens}, \citenamefont {Yard},
  \citenamefont {Yarkoni}, \citenamefont {Zechiel}, \citenamefont {Zhuk},\ and\
  \citenamefont {Zoufal}}]{Abbas2024NatRevPhys}%
  \BibitemOpen
  \bibfield  {author} {\bibinfo {author} {\bibfnamefont {A.}~\bibnamefont
  {Abbas}}, \bibinfo {author} {\bibfnamefont {A.}~\bibnamefont {Ambainis}},
  \bibinfo {author} {\bibfnamefont {B.}~\bibnamefont {Augustino}}, \bibinfo
  {author} {\bibfnamefont {A.}~\bibnamefont {B{\"a}rtschi}}, \bibinfo {author}
  {\bibfnamefont {H.}~\bibnamefont {Buhrman}}, \bibinfo {author} {\bibfnamefont
  {C.}~\bibnamefont {Coffrin}}, \bibinfo {author} {\bibfnamefont
  {G.}~\bibnamefont {Cortiana}}, \bibinfo {author} {\bibfnamefont
  {V.}~\bibnamefont {Dunjko}}, \bibinfo {author} {\bibfnamefont {D.~J.}\
  \bibnamefont {Egger}}, \bibinfo {author} {\bibfnamefont {B.~G.}\ \bibnamefont
  {Elmegreen}}, \bibinfo {author} {\bibfnamefont {N.}~\bibnamefont {Franco}},
  \bibinfo {author} {\bibfnamefont {F.}~\bibnamefont {Fratini}}, \bibinfo
  {author} {\bibfnamefont {B.}~\bibnamefont {Fuller}}, \bibinfo {author}
  {\bibfnamefont {J.}~\bibnamefont {Gacon}}, \bibinfo {author} {\bibfnamefont
  {C.}~\bibnamefont {Gonciulea}}, \bibinfo {author} {\bibfnamefont
  {S.}~\bibnamefont {Gribling}}, \bibinfo {author} {\bibfnamefont
  {S.}~\bibnamefont {Gupta}}, \bibinfo {author} {\bibfnamefont
  {S.}~\bibnamefont {Hadfield}}, \bibinfo {author} {\bibfnamefont
  {R.}~\bibnamefont {Heese}}, \bibinfo {author} {\bibfnamefont
  {G.}~\bibnamefont {Kircher}}, \bibinfo {author} {\bibfnamefont
  {T.}~\bibnamefont {Kleinert}}, \bibinfo {author} {\bibfnamefont
  {T.}~\bibnamefont {Koch}}, \bibinfo {author} {\bibfnamefont {G.}~\bibnamefont
  {Korpas}}, \bibinfo {author} {\bibfnamefont {S.}~\bibnamefont {Lenk}},
  \bibinfo {author} {\bibfnamefont {J.}~\bibnamefont {Marecek}}, \bibinfo
  {author} {\bibfnamefont {V.}~\bibnamefont {Markov}}, \bibinfo {author}
  {\bibfnamefont {G.}~\bibnamefont {Mazzola}}, \bibinfo {author} {\bibfnamefont
  {S.}~\bibnamefont {Mensa}}, \bibinfo {author} {\bibfnamefont
  {N.}~\bibnamefont {Mohseni}}, \bibinfo {author} {\bibfnamefont
  {G.}~\bibnamefont {Nannicini}}, \bibinfo {author} {\bibfnamefont
  {C.}~\bibnamefont {O'Meara}}, \bibinfo {author} {\bibfnamefont {E.~P.}\
  \bibnamefont {Tapia}}, \bibinfo {author} {\bibfnamefont {S.}~\bibnamefont
  {Pokutta}}, \bibinfo {author} {\bibfnamefont {M.}~\bibnamefont {Proissl}},
  \bibinfo {author} {\bibfnamefont {P.}~\bibnamefont {Rebentrost}}, \bibinfo
  {author} {\bibfnamefont {E.}~\bibnamefont {Sahin}}, \bibinfo {author}
  {\bibfnamefont {B.~C.~B.}\ \bibnamefont {Symons}}, \bibinfo {author}
  {\bibfnamefont {S.}~\bibnamefont {Tornow}}, \bibinfo {author} {\bibfnamefont
  {V.}~\bibnamefont {Valls}}, \bibinfo {author} {\bibfnamefont
  {S.}~\bibnamefont {Woerner}}, \bibinfo {author} {\bibfnamefont {M.~L.}\
  \bibnamefont {Wolf-Bauwens}}, \bibinfo {author} {\bibfnamefont
  {J.}~\bibnamefont {Yard}}, \bibinfo {author} {\bibfnamefont {S.}~\bibnamefont
  {Yarkoni}}, \bibinfo {author} {\bibfnamefont {D.}~\bibnamefont {Zechiel}},
  \bibinfo {author} {\bibfnamefont {S.}~\bibnamefont {Zhuk}},\ and\ \bibinfo
  {author} {\bibfnamefont {C.}~\bibnamefont {Zoufal}},\ }\bibfield  {title}
  {\bibinfo {title} {Challenges and opportunities in quantum optimization},\
  }\href {https://doi.org/10.1038/s42254-024-00770-9} {\bibfield  {journal}
  {\bibinfo  {journal} {Nature Reviews Physics}\ }\textbf {\bibinfo {volume}
  {6}},\ \bibinfo {pages} {718} (\bibinfo {year} {2024})}\BibitemShut {NoStop}%
\bibitem [{\citenamefont {Georgescu}\ \emph {et~al.}(2014)\citenamefont
  {Georgescu}, \citenamefont {Ashhab},\ and\ \citenamefont
  {Nori}}]{Georgescu2014RevModPhys}%
  \BibitemOpen
  \bibfield  {author} {\bibinfo {author} {\bibfnamefont {I.~M.}\ \bibnamefont
  {Georgescu}}, \bibinfo {author} {\bibfnamefont {S.}~\bibnamefont {Ashhab}},\
  and\ \bibinfo {author} {\bibfnamefont {F.}~\bibnamefont {Nori}},\ }\bibfield
  {title} {\bibinfo {title} {Quantum simulation},\ }\href
  {https://doi.org/10.1103/RevModPhys.86.153} {\bibfield  {journal} {\bibinfo
  {journal} {Rev. Mod. Phys.}\ }\textbf {\bibinfo {volume} {86}},\ \bibinfo
  {pages} {153} (\bibinfo {year} {2014})}\BibitemShut {NoStop}%
\bibitem [{\citenamefont {Daley}\ \emph {et~al.}(2022)\citenamefont {Daley},
  \citenamefont {Bloch}, \citenamefont {Kokail}, \citenamefont {Flannigan},
  \citenamefont {Pearson}, \citenamefont {Troyer},\ and\ \citenamefont
  {Zoller}}]{Daley2022Nature}%
  \BibitemOpen
  \bibfield  {author} {\bibinfo {author} {\bibfnamefont {A.~J.}\ \bibnamefont
  {Daley}}, \bibinfo {author} {\bibfnamefont {I.}~\bibnamefont {Bloch}},
  \bibinfo {author} {\bibfnamefont {C.}~\bibnamefont {Kokail}}, \bibinfo
  {author} {\bibfnamefont {S.}~\bibnamefont {Flannigan}}, \bibinfo {author}
  {\bibfnamefont {N.}~\bibnamefont {Pearson}}, \bibinfo {author} {\bibfnamefont
  {M.}~\bibnamefont {Troyer}},\ and\ \bibinfo {author} {\bibfnamefont
  {P.}~\bibnamefont {Zoller}},\ }\bibfield  {title} {\bibinfo {title}
  {Practical quantum advantage in quantum simulation},\ }\href
  {https://doi.org/10.1038/s41586-022-04940-6} {\bibfield  {journal} {\bibinfo
  {journal} {Nature}\ }\textbf {\bibinfo {volume} {607}},\ \bibinfo {pages}
  {667} (\bibinfo {year} {2022})}\BibitemShut {NoStop}%
\bibitem [{\citenamefont {Barends}\ \emph {et~al.}(2016)\citenamefont
  {Barends}, \citenamefont {Shabani}, \citenamefont {Lamata}, \citenamefont
  {Kelly}, \citenamefont {Mezzacapo}, \citenamefont {Heras}, \citenamefont
  {Babbush}, \citenamefont {Fowler}, \citenamefont {Campbell}, \citenamefont
  {Chen}, \citenamefont {Chen}, \citenamefont {Chiaro}, \citenamefont
  {Dunsworth}, \citenamefont {Jeffrey}, \citenamefont {Lucero}, \citenamefont
  {Megrant}, \citenamefont {Mutus}, \citenamefont {Neeley}, \citenamefont
  {Neill}, \citenamefont {O'Malley}, \citenamefont {Quintana}, \citenamefont
  {Roushan}, \citenamefont {Sank}, \citenamefont {Vainsencher}, \citenamefont
  {Wenner}, \citenamefont {White}, \citenamefont {Solano}, \citenamefont
  {Neven},\ and\ \citenamefont {Martinis}}]{Barends2016Nature}%
  \BibitemOpen
  \bibfield  {author} {\bibinfo {author} {\bibfnamefont {R.}~\bibnamefont
  {Barends}}, \bibinfo {author} {\bibfnamefont {A.}~\bibnamefont {Shabani}},
  \bibinfo {author} {\bibfnamefont {L.}~\bibnamefont {Lamata}}, \bibinfo
  {author} {\bibfnamefont {J.}~\bibnamefont {Kelly}}, \bibinfo {author}
  {\bibfnamefont {A.}~\bibnamefont {Mezzacapo}}, \bibinfo {author}
  {\bibfnamefont {U.~L.}\ \bibnamefont {Heras}}, \bibinfo {author}
  {\bibfnamefont {R.}~\bibnamefont {Babbush}}, \bibinfo {author} {\bibfnamefont
  {A.~G.}\ \bibnamefont {Fowler}}, \bibinfo {author} {\bibfnamefont
  {B.}~\bibnamefont {Campbell}}, \bibinfo {author} {\bibfnamefont
  {Y.}~\bibnamefont {Chen}}, \bibinfo {author} {\bibfnamefont {Z.}~\bibnamefont
  {Chen}}, \bibinfo {author} {\bibfnamefont {B.}~\bibnamefont {Chiaro}},
  \bibinfo {author} {\bibfnamefont {A.}~\bibnamefont {Dunsworth}}, \bibinfo
  {author} {\bibfnamefont {E.}~\bibnamefont {Jeffrey}}, \bibinfo {author}
  {\bibfnamefont {E.}~\bibnamefont {Lucero}}, \bibinfo {author} {\bibfnamefont
  {A.}~\bibnamefont {Megrant}}, \bibinfo {author} {\bibfnamefont {J.~Y.}\
  \bibnamefont {Mutus}}, \bibinfo {author} {\bibfnamefont {M.}~\bibnamefont
  {Neeley}}, \bibinfo {author} {\bibfnamefont {C.}~\bibnamefont {Neill}},
  \bibinfo {author} {\bibfnamefont {P.~J.~J.}\ \bibnamefont {O'Malley}},
  \bibinfo {author} {\bibfnamefont {C.}~\bibnamefont {Quintana}}, \bibinfo
  {author} {\bibfnamefont {P.}~\bibnamefont {Roushan}}, \bibinfo {author}
  {\bibfnamefont {D.}~\bibnamefont {Sank}}, \bibinfo {author} {\bibfnamefont
  {A.}~\bibnamefont {Vainsencher}}, \bibinfo {author} {\bibfnamefont
  {J.}~\bibnamefont {Wenner}}, \bibinfo {author} {\bibfnamefont {T.~C.}\
  \bibnamefont {White}}, \bibinfo {author} {\bibfnamefont {E.}~\bibnamefont
  {Solano}}, \bibinfo {author} {\bibfnamefont {H.}~\bibnamefont {Neven}},\ and\
  \bibinfo {author} {\bibfnamefont {J.~M.}\ \bibnamefont {Martinis}},\
  }\bibfield  {title} {\bibinfo {title} {Digitized adiabatic quantum computing
  with a superconducting circuit},\ }\href
  {https://doi.org/10.1038/nature17658} {\bibfield  {journal} {\bibinfo
  {journal} {Nature}\ }\textbf {\bibinfo {volume} {534}},\ \bibinfo {pages}
  {222} (\bibinfo {year} {2016})}\BibitemShut {NoStop}%
\bibitem [{\citenamefont {Albash}\ and\ \citenamefont
  {Lidar}(2018)}]{Albash2018RevModPhys}%
  \BibitemOpen
  \bibfield  {author} {\bibinfo {author} {\bibfnamefont {T.}~\bibnamefont
  {Albash}}\ and\ \bibinfo {author} {\bibfnamefont {D.~A.}\ \bibnamefont
  {Lidar}},\ }\bibfield  {title} {\bibinfo {title} {Adiabatic quantum
  computation},\ }\href {https://doi.org/10.1103/RevModPhys.90.015002}
  {\bibfield  {journal} {\bibinfo  {journal} {Rev. Mod. Phys.}\ }\textbf
  {\bibinfo {volume} {90}},\ \bibinfo {pages} {015002} (\bibinfo {year}
  {2018})}\BibitemShut {NoStop}%
\bibitem [{\citenamefont {Garc\'{\i}a-Pintos}\ \emph
  {et~al.}(2023)\citenamefont {Garc\'{\i}a-Pintos}, \citenamefont {Brady},
  \citenamefont {Bringewatt},\ and\ \citenamefont
  {Liu}}]{GarciaPintos2023PhysRevLett}%
  \BibitemOpen
  \bibfield  {author} {\bibinfo {author} {\bibfnamefont {L.~P.}\ \bibnamefont
  {Garc\'{\i}a-Pintos}}, \bibinfo {author} {\bibfnamefont {L.~T.}\ \bibnamefont
  {Brady}}, \bibinfo {author} {\bibfnamefont {J.}~\bibnamefont {Bringewatt}},\
  and\ \bibinfo {author} {\bibfnamefont {Y.-K.}\ \bibnamefont {Liu}},\
  }\bibfield  {title} {\bibinfo {title} {Lower bounds on quantum annealing
  times},\ }\href {https://doi.org/10.1103/PhysRevLett.130.140601} {\bibfield
  {journal} {\bibinfo  {journal} {Phys. Rev. Lett.}\ }\textbf {\bibinfo
  {volume} {130}},\ \bibinfo {pages} {140601} (\bibinfo {year}
  {2023})}\BibitemShut {NoStop}%
\bibitem [{\citenamefont {Chen}\ \emph {et~al.}(2010)\citenamefont {Chen},
  \citenamefont {Ruschhaupt}, \citenamefont {Schmidt}, \citenamefont {del
  Campo}, \citenamefont {Gu\'ery-Odelin},\ and\ \citenamefont
  {Muga}}]{Chen2010PhysRevLett}%
  \BibitemOpen
  \bibfield  {author} {\bibinfo {author} {\bibfnamefont {X.}~\bibnamefont
  {Chen}}, \bibinfo {author} {\bibfnamefont {A.}~\bibnamefont {Ruschhaupt}},
  \bibinfo {author} {\bibfnamefont {S.}~\bibnamefont {Schmidt}}, \bibinfo
  {author} {\bibfnamefont {A.}~\bibnamefont {del Campo}}, \bibinfo {author}
  {\bibfnamefont {D.}~\bibnamefont {Gu\'ery-Odelin}},\ and\ \bibinfo {author}
  {\bibfnamefont {J.~G.}\ \bibnamefont {Muga}},\ }\bibfield  {title} {\bibinfo
  {title} {Fast optimal frictionless atom cooling in harmonic traps: Shortcut
  to adiabaticity},\ }\href {https://doi.org/10.1103/PhysRevLett.104.063002}
  {\bibfield  {journal} {\bibinfo  {journal} {Phys. Rev. Lett.}\ }\textbf
  {\bibinfo {volume} {104}},\ \bibinfo {pages} {063002} (\bibinfo {year}
  {2010})}\BibitemShut {NoStop}%
\bibitem [{\citenamefont {Gu\'ery-Odelin}\ \emph {et~al.}(2019)\citenamefont
  {Gu\'ery-Odelin}, \citenamefont {Ruschhaupt}, \citenamefont {Kiely},
  \citenamefont {Torrontegui}, \citenamefont {Mart\'{\i}nez-Garaot},\ and\
  \citenamefont {Muga}}]{GueryOdelin2019RevModPhys}%
  \BibitemOpen
  \bibfield  {author} {\bibinfo {author} {\bibfnamefont {D.}~\bibnamefont
  {Gu\'ery-Odelin}}, \bibinfo {author} {\bibfnamefont {A.}~\bibnamefont
  {Ruschhaupt}}, \bibinfo {author} {\bibfnamefont {A.}~\bibnamefont {Kiely}},
  \bibinfo {author} {\bibfnamefont {E.}~\bibnamefont {Torrontegui}}, \bibinfo
  {author} {\bibfnamefont {S.}~\bibnamefont {Mart\'{\i}nez-Garaot}},\ and\
  \bibinfo {author} {\bibfnamefont {J.~G.}\ \bibnamefont {Muga}},\ }\bibfield
  {title} {\bibinfo {title} {Shortcuts to adiabaticity: Concepts, methods, and
  applications},\ }\href {https://doi.org/10.1103/RevModPhys.91.045001}
  {\bibfield  {journal} {\bibinfo  {journal} {Rev. Mod. Phys.}\ }\textbf
  {\bibinfo {volume} {91}},\ \bibinfo {pages} {045001} (\bibinfo {year}
  {2019})}\BibitemShut {NoStop}%
\bibitem [{\citenamefont {del Campo}(2013)}]{DelCampo2013PhysRevLett}%
  \BibitemOpen
  \bibfield  {author} {\bibinfo {author} {\bibfnamefont {A.}~\bibnamefont {del
  Campo}},\ }\bibfield  {title} {\bibinfo {title} {Shortcuts to adiabaticity by
  counterdiabatic driving},\ }\href
  {https://doi.org/10.1103/PhysRevLett.111.100502} {\bibfield  {journal}
  {\bibinfo  {journal} {Phys. Rev. Lett.}\ }\textbf {\bibinfo {volume} {111}},\
  \bibinfo {pages} {100502} (\bibinfo {year} {2013})}\BibitemShut {NoStop}%
\bibitem [{\citenamefont {Nakahara}(2022)}]{Nakahara2022PTRSA}%
  \BibitemOpen
  \bibfield  {author} {\bibinfo {author} {\bibfnamefont {M.}~\bibnamefont
  {Nakahara}},\ }\bibfield  {title} {\bibinfo {title} {Counterdiabatic
  formalism of shortcuts to adiabaticity},\ }\href
  {https://doi.org/10.1098/rsta.2021.0272} {\bibfield  {journal} {\bibinfo
  {journal} {Philosophical Transactions of the Royal Society A: Mathematical,
  Physical and Engineering Sciences}\ }\textbf {\bibinfo {volume} {380}},\
  \bibinfo {pages} {20210272} (\bibinfo {year} {2022})},\ \Eprint
  {https://arxiv.org/abs/https://royalsocietypublishing.org/rsta/article-pdf/doi/10.1098/rsta.2021.0272/1327084/rsta.2021.0272.pdf}
  {https://royalsocietypublishing.org/rsta/article-pdf/doi/10.1098/rsta.2021.0272/1327084/rsta.2021.0272.pdf}
  \BibitemShut {NoStop}%
\bibitem [{\citenamefont {Hegade}\ \emph {et~al.}(2021)\citenamefont {Hegade},
  \citenamefont {Paul}, \citenamefont {Ding}, \citenamefont {Sanz},
  \citenamefont {Albarr\'an-Arriagada}, \citenamefont {Solano},\ and\
  \citenamefont {Chen}}]{Hegade2021PhysRevAppl}%
  \BibitemOpen
  \bibfield  {author} {\bibinfo {author} {\bibfnamefont {N.~N.}\ \bibnamefont
  {Hegade}}, \bibinfo {author} {\bibfnamefont {K.}~\bibnamefont {Paul}},
  \bibinfo {author} {\bibfnamefont {Y.}~\bibnamefont {Ding}}, \bibinfo {author}
  {\bibfnamefont {M.}~\bibnamefont {Sanz}}, \bibinfo {author} {\bibfnamefont
  {F.}~\bibnamefont {Albarr\'an-Arriagada}}, \bibinfo {author} {\bibfnamefont
  {E.}~\bibnamefont {Solano}},\ and\ \bibinfo {author} {\bibfnamefont
  {X.}~\bibnamefont {Chen}},\ }\bibfield  {title} {\bibinfo {title} {Shortcuts
  to adiabaticity in digitized adiabatic quantum computing},\ }\href
  {https://doi.org/10.1103/PhysRevApplied.15.024038} {\bibfield  {journal}
  {\bibinfo  {journal} {Phys. Rev. Appl.}\ }\textbf {\bibinfo {volume} {15}},\
  \bibinfo {pages} {024038} (\bibinfo {year} {2021})}\BibitemShut {NoStop}%
\bibitem [{\citenamefont {Wang}\ \emph {et~al.}(2019)\citenamefont {Wang},
  \citenamefont {Zhang}, \citenamefont {Xiang}, \citenamefont {Jia},
  \citenamefont {Duan}, \citenamefont {Zong}, \citenamefont {Sun},
  \citenamefont {Dong}, \citenamefont {Wu}, \citenamefont {Yin},\ and\
  \citenamefont {Guo}}]{Wang2019PhysRevAppl}%
  \BibitemOpen
  \bibfield  {author} {\bibinfo {author} {\bibfnamefont {T.}~\bibnamefont
  {Wang}}, \bibinfo {author} {\bibfnamefont {Z.}~\bibnamefont {Zhang}},
  \bibinfo {author} {\bibfnamefont {L.}~\bibnamefont {Xiang}}, \bibinfo
  {author} {\bibfnamefont {Z.}~\bibnamefont {Jia}}, \bibinfo {author}
  {\bibfnamefont {P.}~\bibnamefont {Duan}}, \bibinfo {author} {\bibfnamefont
  {Z.}~\bibnamefont {Zong}}, \bibinfo {author} {\bibfnamefont {Z.}~\bibnamefont
  {Sun}}, \bibinfo {author} {\bibfnamefont {Z.}~\bibnamefont {Dong}}, \bibinfo
  {author} {\bibfnamefont {J.}~\bibnamefont {Wu}}, \bibinfo {author}
  {\bibfnamefont {Y.}~\bibnamefont {Yin}},\ and\ \bibinfo {author}
  {\bibfnamefont {G.}~\bibnamefont {Guo}},\ }\bibfield  {title} {\bibinfo
  {title} {Experimental realization of a fast controlled-z gate via a shortcut
  to adiabaticity},\ }\href {https://doi.org/10.1103/PhysRevApplied.11.034030}
  {\bibfield  {journal} {\bibinfo  {journal} {Phys. Rev. Appl.}\ }\textbf
  {\bibinfo {volume} {11}},\ \bibinfo {pages} {034030} (\bibinfo {year}
  {2019})}\BibitemShut {NoStop}%
\bibitem [{\citenamefont {Zhou}\ \emph {et~al.}(2020)\citenamefont {Zhou},
  \citenamefont {Ji}, \citenamefont {Nie}, \citenamefont {Yang}, \citenamefont
  {Chen}, \citenamefont {Bian},\ and\ \citenamefont
  {Peng}}]{Zhou2020PhysRevAppl}%
  \BibitemOpen
  \bibfield  {author} {\bibinfo {author} {\bibfnamefont {H.}~\bibnamefont
  {Zhou}}, \bibinfo {author} {\bibfnamefont {Y.}~\bibnamefont {Ji}}, \bibinfo
  {author} {\bibfnamefont {X.}~\bibnamefont {Nie}}, \bibinfo {author}
  {\bibfnamefont {X.}~\bibnamefont {Yang}}, \bibinfo {author} {\bibfnamefont
  {X.}~\bibnamefont {Chen}}, \bibinfo {author} {\bibfnamefont {J.}~\bibnamefont
  {Bian}},\ and\ \bibinfo {author} {\bibfnamefont {X.}~\bibnamefont {Peng}},\
  }\bibfield  {title} {\bibinfo {title} {Experimental realization of shortcuts
  to adiabaticity in a nonintegrable spin chain by local counterdiabatic
  driving},\ }\href {https://doi.org/10.1103/PhysRevApplied.13.044059}
  {\bibfield  {journal} {\bibinfo  {journal} {Phys. Rev. Appl.}\ }\textbf
  {\bibinfo {volume} {13}},\ \bibinfo {pages} {044059} (\bibinfo {year}
  {2020})}\BibitemShut {NoStop}%
\bibitem [{\citenamefont {Zhang}\ \emph {et~al.}(2024)\citenamefont {Zhang},
  \citenamefont {Hegade}, \citenamefont {Cadavid}, \citenamefont
  {Lassabli{\`e}re}, \citenamefont {Trautmann}, \citenamefont {Perseguers},
  \citenamefont {Solano}, \citenamefont {Henriet},\ and\ \citenamefont
  {Michon}}]{Zhang2024arXiv}%
  \BibitemOpen
  \bibfield  {author} {\bibinfo {author} {\bibfnamefont {Q.}~\bibnamefont
  {Zhang}}, \bibinfo {author} {\bibfnamefont {N.~N.}\ \bibnamefont {Hegade}},
  \bibinfo {author} {\bibfnamefont {A.~G.}\ \bibnamefont {Cadavid}}, \bibinfo
  {author} {\bibfnamefont {L.}~\bibnamefont {Lassabli{\`e}re}}, \bibinfo
  {author} {\bibfnamefont {J.}~\bibnamefont {Trautmann}}, \bibinfo {author}
  {\bibfnamefont {S.}~\bibnamefont {Perseguers}}, \bibinfo {author}
  {\bibfnamefont {E.}~\bibnamefont {Solano}}, \bibinfo {author} {\bibfnamefont
  {L.}~\bibnamefont {Henriet}},\ and\ \bibinfo {author} {\bibfnamefont
  {E.}~\bibnamefont {Michon}},\ }\bibfield  {title} {\bibinfo {title} {Analog
  counterdiabatic quantum computing},\ }\href@noop {} {\bibfield  {journal}
  {\bibinfo  {journal} {arXiv preprint arXiv:2405.14829}\ } (\bibinfo {year}
  {2024})}\BibitemShut {NoStop}%
\bibitem [{\citenamefont {Masuda}\ and\ \citenamefont
  {Nakamura}(2022)}]{Masuda2022PTRSA}%
  \BibitemOpen
  \bibfield  {author} {\bibinfo {author} {\bibfnamefont {S.}~\bibnamefont
  {Masuda}}\ and\ \bibinfo {author} {\bibfnamefont {K.}~\bibnamefont
  {Nakamura}},\ }\bibfield  {title} {\bibinfo {title} {Fast-forward scaling
  theory},\ }\href {https://doi.org/10.1098/rsta.2021.0278} {\bibfield
  {journal} {\bibinfo  {journal} {Philosophical Transactions of the Royal
  Society A: Mathematical, Physical and Engineering Sciences}\ }\textbf
  {\bibinfo {volume} {380}},\ \bibinfo {pages} {20210278} (\bibinfo {year}
  {2022})},\ \Eprint
  {https://arxiv.org/abs/https://royalsocietypublishing.org/rsta/article-pdf/doi/10.1098/rsta.2021.0278/1326805/rsta.2021.0278.pdf}
  {https://royalsocietypublishing.org/rsta/article-pdf/doi/10.1098/rsta.2021.0278/1326805/rsta.2021.0278.pdf}
  \BibitemShut {NoStop}%
\bibitem [{\citenamefont {Bernardo}(2020)}]{Bernardo2020PhysRevRes}%
  \BibitemOpen
  \bibfield  {author} {\bibinfo {author} {\bibfnamefont {B.~d.~L.}\
  \bibnamefont {Bernardo}},\ }\bibfield  {title} {\bibinfo {title}
  {Time-rescaled quantum dynamics as a shortcut to adiabaticity},\ }\href
  {https://doi.org/10.1103/PhysRevResearch.2.013133} {\bibfield  {journal}
  {\bibinfo  {journal} {Phys. Rev. Res.}\ }\textbf {\bibinfo {volume} {2}},\
  \bibinfo {pages} {013133} (\bibinfo {year} {2020})}\BibitemShut {NoStop}%
\bibitem [{\citenamefont {Sels}\ and\ \citenamefont
  {Polkovnikov}(2017)}]{Sels2017PNAS}%
  \BibitemOpen
  \bibfield  {author} {\bibinfo {author} {\bibfnamefont {D.}~\bibnamefont
  {Sels}}\ and\ \bibinfo {author} {\bibfnamefont {A.}~\bibnamefont
  {Polkovnikov}},\ }\bibfield  {title} {\bibinfo {title} {Minimizing
  irreversible losses in quantum systems by local counterdiabatic driving},\
  }\href {https://doi.org/10.1073/pnas.1619826114} {\bibfield  {journal}
  {\bibinfo  {journal} {Proceedings of the National Academy of Sciences}\
  }\textbf {\bibinfo {volume} {114}},\ \bibinfo {pages} {E3909} (\bibinfo
  {year} {2017})}\BibitemShut {NoStop}%
\bibitem [{\citenamefont {Fin\ifmmode~\check{z}\else \v{z}\fi{}gar}\ \emph
  {et~al.}(2024)\citenamefont {Fin\ifmmode~\check{z}\else \v{z}\fi{}gar},
  \citenamefont {Schuetz}, \citenamefont {Brubaker}, \citenamefont
  {Nishimori},\ and\ \citenamefont {Katzgraber}}]{Finzgar2024PhysRevRes}%
  \BibitemOpen
  \bibfield  {author} {\bibinfo {author} {\bibfnamefont {J.~R.}\ \bibnamefont
  {Fin\ifmmode~\check{z}\else \v{z}\fi{}gar}}, \bibinfo {author} {\bibfnamefont
  {M.~J.~A.}\ \bibnamefont {Schuetz}}, \bibinfo {author} {\bibfnamefont
  {J.~K.}\ \bibnamefont {Brubaker}}, \bibinfo {author} {\bibfnamefont
  {H.}~\bibnamefont {Nishimori}},\ and\ \bibinfo {author} {\bibfnamefont
  {H.~G.}\ \bibnamefont {Katzgraber}},\ }\bibfield  {title} {\bibinfo {title}
  {Designing quantum annealing schedules using bayesian optimization},\ }\href
  {https://doi.org/10.1103/PhysRevResearch.6.023063} {\bibfield  {journal}
  {\bibinfo  {journal} {Phys. Rev. Res.}\ }\textbf {\bibinfo {volume} {6}},\
  \bibinfo {pages} {023063} (\bibinfo {year} {2024})}\BibitemShut {NoStop}%
\bibitem [{\citenamefont {Gu}\ \emph {et~al.}(2003)\citenamefont {Gu},
  \citenamefont {Lin},\ and\ \citenamefont {Li}}]{Gu2003PhysRevA}%
  \BibitemOpen
  \bibfield  {author} {\bibinfo {author} {\bibfnamefont {S.-J.}\ \bibnamefont
  {Gu}}, \bibinfo {author} {\bibfnamefont {H.-Q.}\ \bibnamefont {Lin}},\ and\
  \bibinfo {author} {\bibfnamefont {Y.-Q.}\ \bibnamefont {Li}},\ }\bibfield
  {title} {\bibinfo {title} {Entanglement, quantum phase transition, and
  scaling in the $\mathrm{XXZ}$ chain},\ }\href
  {https://doi.org/10.1103/PhysRevA.68.042330} {\bibfield  {journal} {\bibinfo
  {journal} {Phys. Rev. A}\ }\textbf {\bibinfo {volume} {68}},\ \bibinfo
  {pages} {042330} (\bibinfo {year} {2003})}\BibitemShut {NoStop}%
\bibitem [{\citenamefont {Gu}\ \emph {et~al.}(2005)\citenamefont {Gu},
  \citenamefont {Tian},\ and\ \citenamefont {Lin}}]{Gu2005PhysRevA}%
  \BibitemOpen
  \bibfield  {author} {\bibinfo {author} {\bibfnamefont {S.-J.}\ \bibnamefont
  {Gu}}, \bibinfo {author} {\bibfnamefont {G.-S.}\ \bibnamefont {Tian}},\ and\
  \bibinfo {author} {\bibfnamefont {H.-Q.}\ \bibnamefont {Lin}},\ }\bibfield
  {title} {\bibinfo {title} {Ground-state entanglement in the $xxz$ model},\
  }\href {https://doi.org/10.1103/PhysRevA.71.052322} {\bibfield  {journal}
  {\bibinfo  {journal} {Phys. Rev. A}\ }\textbf {\bibinfo {volume} {71}},\
  \bibinfo {pages} {052322} (\bibinfo {year} {2005})}\BibitemShut {NoStop}%
\bibitem [{\citenamefont {Kubo}\ and\ \citenamefont
  {Kishi}(1988)}]{Kubo1988PhysRevLett}%
  \BibitemOpen
  \bibfield  {author} {\bibinfo {author} {\bibfnamefont {K.}~\bibnamefont
  {Kubo}}\ and\ \bibinfo {author} {\bibfnamefont {T.}~\bibnamefont {Kishi}},\
  }\bibfield  {title} {\bibinfo {title} {Existence of long-range order in the
  $\mathrm{XXZ}$ model},\ }\href {https://doi.org/10.1103/PhysRevLett.61.2585}
  {\bibfield  {journal} {\bibinfo  {journal} {Phys. Rev. Lett.}\ }\textbf
  {\bibinfo {volume} {61}},\ \bibinfo {pages} {2585} (\bibinfo {year}
  {1988})}\BibitemShut {NoStop}%
\bibitem [{\citenamefont {Berry}(2009)}]{Berry2009JPhysA}%
  \BibitemOpen
  \bibfield  {author} {\bibinfo {author} {\bibfnamefont {M.~V.}\ \bibnamefont
  {Berry}},\ }\bibfield  {title} {\bibinfo {title} {Transitionless quantum
  driving},\ }\href {https://doi.org/10.1088/1751-8113/42/36/365303} {\bibfield
   {journal} {\bibinfo  {journal} {Journal of Physics A: Mathematical and
  Theoretical}\ }\textbf {\bibinfo {volume} {42}},\ \bibinfo {pages} {365303}
  (\bibinfo {year} {2009})}\BibitemShut {NoStop}%
\bibitem [{\citenamefont {Claeys}\ \emph {et~al.}(2019)\citenamefont {Claeys},
  \citenamefont {Pandey}, \citenamefont {Sels},\ and\ \citenamefont
  {Polkovnikov}}]{Claeys2019PhysRevLett}%
  \BibitemOpen
  \bibfield  {author} {\bibinfo {author} {\bibfnamefont {P.~W.}\ \bibnamefont
  {Claeys}}, \bibinfo {author} {\bibfnamefont {M.}~\bibnamefont {Pandey}},
  \bibinfo {author} {\bibfnamefont {D.}~\bibnamefont {Sels}},\ and\ \bibinfo
  {author} {\bibfnamefont {A.}~\bibnamefont {Polkovnikov}},\ }\bibfield
  {title} {\bibinfo {title} {Floquet-engineering counterdiabatic protocols in
  quantum many-body systems},\ }\href
  {https://doi.org/10.1103/PhysRevLett.123.090602} {\bibfield  {journal}
  {\bibinfo  {journal} {Phys. Rev. Lett.}\ }\textbf {\bibinfo {volume} {123}},\
  \bibinfo {pages} {090602} (\bibinfo {year} {2019})}\BibitemShut {NoStop}%
\bibitem [{\citenamefont {Xie}\ \emph {et~al.}(2022)\citenamefont {Xie},
  \citenamefont {Seki},\ and\ \citenamefont {Yunoki}}]{Xie2022PhysRevB}%
  \BibitemOpen
  \bibfield  {author} {\bibinfo {author} {\bibfnamefont {Q.}~\bibnamefont
  {Xie}}, \bibinfo {author} {\bibfnamefont {K.}~\bibnamefont {Seki}},\ and\
  \bibinfo {author} {\bibfnamefont {S.}~\bibnamefont {Yunoki}},\ }\bibfield
  {title} {\bibinfo {title} {Variational counterdiabatic driving of the hubbard
  model for ground-state preparation},\ }\href
  {https://doi.org/10.1103/PhysRevB.106.155153} {\bibfield  {journal} {\bibinfo
   {journal} {Phys. Rev. B}\ }\textbf {\bibinfo {volume} {106}},\ \bibinfo
  {pages} {155153} (\bibinfo {year} {2022})}\BibitemShut {NoStop}%
\end{thebibliography}%

\end{document}